# Effects of Grain Boundary Disorder on Yield Strength

Valery Borovikov[1], Mikhail I. Mendelev[1] and Alexander H. King[1,2]

[1]*Division of Materials Sciences and Engineering, Ames Laboratory, Ames, IA 50011*
[2]*Department of Materials Science and Engineering, Iowa State University, Ames, IA 50011*

It was recently reported that segregation of Zr to grain boundaries (GB) in nanocrystalline Cu can lead to the formation of disordered intergranular films [1,2]. In this study we employ atomistic computer simulations to study how the formation of these films affects the dislocation nucleation from the GBs. We found that full disorder of the grain boundary structure leads to the suppression of dislocation emission and significant increase of the yield stress. Depending on the solute concentration and heat-treatment, however, a partial disorder may also occur and this aids dislocation nucleation rather than suppressing it, resulting in elimination of the strengthening effect.

Keywords: solute segregation at grain boundaries, dislocation nucleation, yield stress, Monte Carlo simulation, molecular dynamics simulation.

The addition of low-solubility elements is an attractive way to stabilize nanocrystalline materials against grain growth, and Zr is effective at stabilizing nanocrystalline Cu at elevated temperatures because it segregates strongly to the grain boundaries (GB) [3]. Zr segregation at GBs in nanocrystalline Cu, in combination with annealing at elevated temperatures, can lead to the formation of amorphous intergranular films (AIFs) [1,2]. These AIFs provide enhanced thermal stability and lead to improved tensile strength and ductility [1,2,4]. The observed increase in ductility has been explained by the apparent ability of the AIFs to absorb large numbers of incoming dislocations, delaying crack nucleation [5].

It was also recently demonstrated that solute segregation can dramatically impact a GB's ability to emit dislocations in response to stress [6], and because the emission of dislocations from GBs is a key deformation mechanism for many nano/poly-crystalline materials, a solute that stabilizes the microstructure can also increase the yield strength of the material [7,8]. In the present Letter, we describe a set of molecular dynamics (MD) simulations which demonstrate that the effect of Zr on dislocation nucleation in Cu depends on the nature of the disorder at the grain boundary, which is determined by the thermal history of the sample.

In order to perform an MD simulation a suitable interatomic interaction is required. The use of *ab initio* calculations would be the most physically rigorous approach, but this is computationally expensive and cannot be utilized today to simulate plastic deformation over long time-scales in a simulation cell containing tens of thousands of atoms. Therefore, a semi-empirical potential is used, and we have employed a Finnis-Sinclair [9] Cu-Zr potential developed in [10] and available in [11]. A simple bi-crystal geometry, with two $\Sigma 11(332)[110]$ symmetric tilt grain boundaries (STGB) was utilized (see Fig. 1a). This STGB contains E structural units [12], which are known to serve as the dislocation nucleation sources under applied stress [13,14]. All simulations were performed using LAMMPS simulation package [15] and the visualization of the simulation snapshots was performed using the software package OVITO [16]. The simulation cell size was sufficiently large in all three directions (greater than 16 nm in the directions parallel to GB plane, and greater than 32 nm in the direction normal to GB plane) in order to minimize the effect of periodic boundary conditions on the dislocation nucleation [13,17]. The distribution of Zr atoms at the GBs was equilibrated at 300 K using the hybrid Monte Carlo/ molecular dynamics technique described in [18]. Further details of the preparation of the initial models can be found in [6] and the discussion of the Zr segregation pattern can be found in [10]. Additional models were prepared by annealing the initial models described above at 900 K for 10 ns, followed by rapid quenching to 300 K and equilibrating at this temperature. Our simulations reflect the experimental processing applied to the same system in [1,2].

Tensile loading simulations were carried out at 300 K, with a constant engineering strain rate of $10^8$ s$^{-1}$. The deformation was applied in the z direction (normal to the grain boundary plane), while the stresses in the other two directions were held at zero. At the onset of the deformation the stress gradually increases with increasing applied strain, reaches a peak and then drops when the first dislocation is emitted from the grain boundary. This peak stress value is treated as the yield stress in the present study. Figure 1b shows the effects of solute concentration on the yield stress, for unannealed and annealed-and-quenched specimens. In both cases the addition of small amounts of Zr leads to an increase in the yield stress. This effect has been considered and explained in [10], so in the present paper we focus on the effect of the Zr solutes on the yield stress at larger concentrations. For the unannealed models, the yield stress reaches a peak near ~0.83%

Zr concentration and then decreases with further increases of the Zr concentration, and the models with the largest Zr content have almost the same yield stress as the pure Cu. This contrasts with the annealed-and-quenched specimens where Zr additions lead to significant increases in the yield stress in the concentration range from ~0.3 % to ~1 %, beyond which the effect of increasing Zr concentration is not significant.

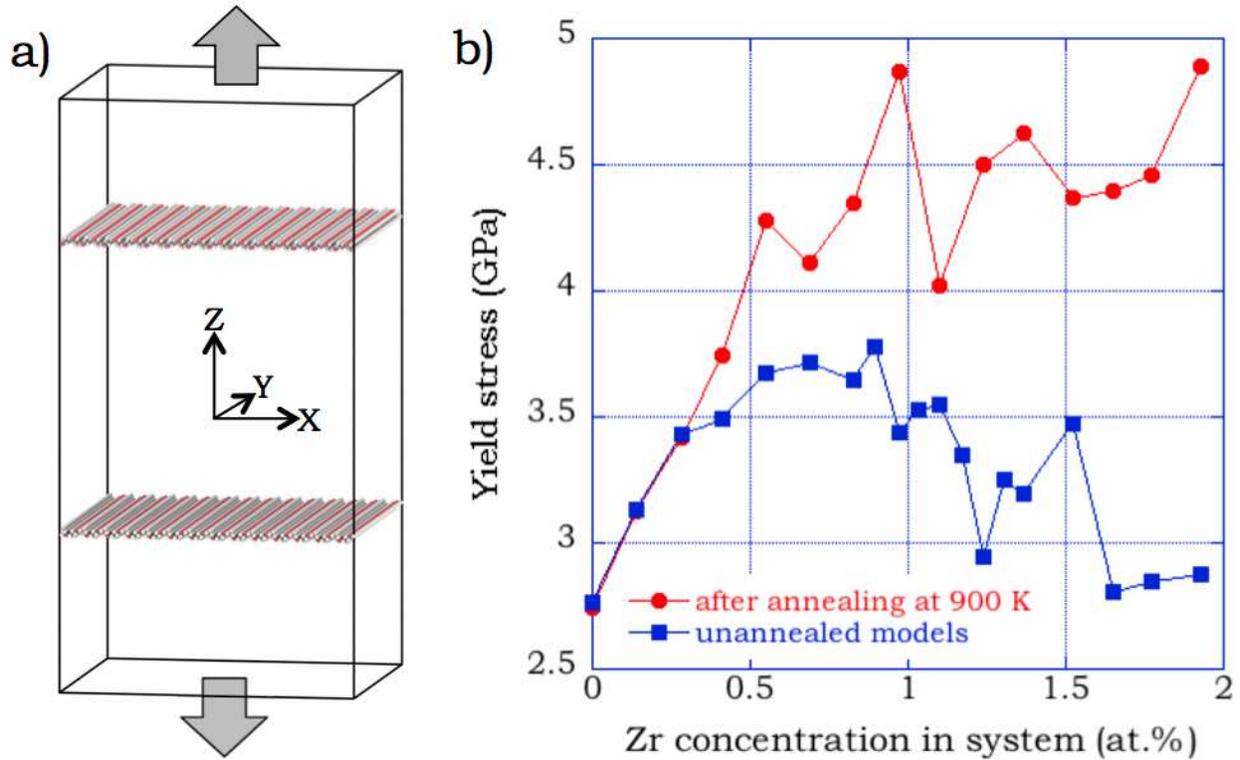

Figure 1. a) Cartoon of the simulation cell. b) The yield stress as function of the solute concentration.

Changes in the GB structure associated with the Zr segregation and the annealing treatment provide an explanation for our results. Annealing does not change the GB structure in pure Cu (see Fig. 2) but can change it in the presence of Zr, so the difference between the yield stresses obtained for the initial and annealed models can be attributed only to the presence of the Zr solutes. The addition of a small amount of Zr (up to 0.28 %) does not significantly change the GB structure even after the annealing, and the MD simulations show almost the same yield stresses for unannealed and annealed specimens in this concentration range.

For the unannealed model, the presence of solutes considerably modifies the GB structure when the Zr concentration is greater than 0.83 % (see Fig. 2). The boundary still retains some E units where the dislocations can nucleate, but these are surrounded by disordered (amorphous) regions, in a structure that corresponds to the description originally proposed by Mott to explain the kinetics of grain boundary sliding and migration [19]. This structure makes dislocation nucleation easier because the atoms in the amorphous regions around the E unit can readily rearrange to accommodate the atom shifts associated with the emission of a dislocation. To illustrate this point we compare the atomic displacements, which occur upon dislocation emission in the cases of lower (0.28%) and higher (1.36%) Zr concentrations. In both cases, we calculated displacements as the changes in the atomic coordinates during the time interval of 22 ps which covers the emission event. Figure 3 shows that for the lower Zr concentration, dislocation emission does not lead to significant atomic reshuffling in the GB region because the ordered, periodic GB structure (seen in the left bottom image in Fig. 3) does not allow such reshuffling. For higher Zr concentration: the GB structure is partially disordered (see the right bottom image in Fig. 3); dislocation nucleation is easily accommodated by atomic reshuffling and requires a smaller applied stress. This explains the decrease of the yield stress at larger Zr concentrations. Figure 2 shows that further increases of the Zr concentration lead to increasing amorphization of the GB structure, but some E units are retained as ordered regions, and dislocation nucleation becomes easier, with the increasing amorphous volume, leading to the decrease in the yield stress (see Fig. 1b).

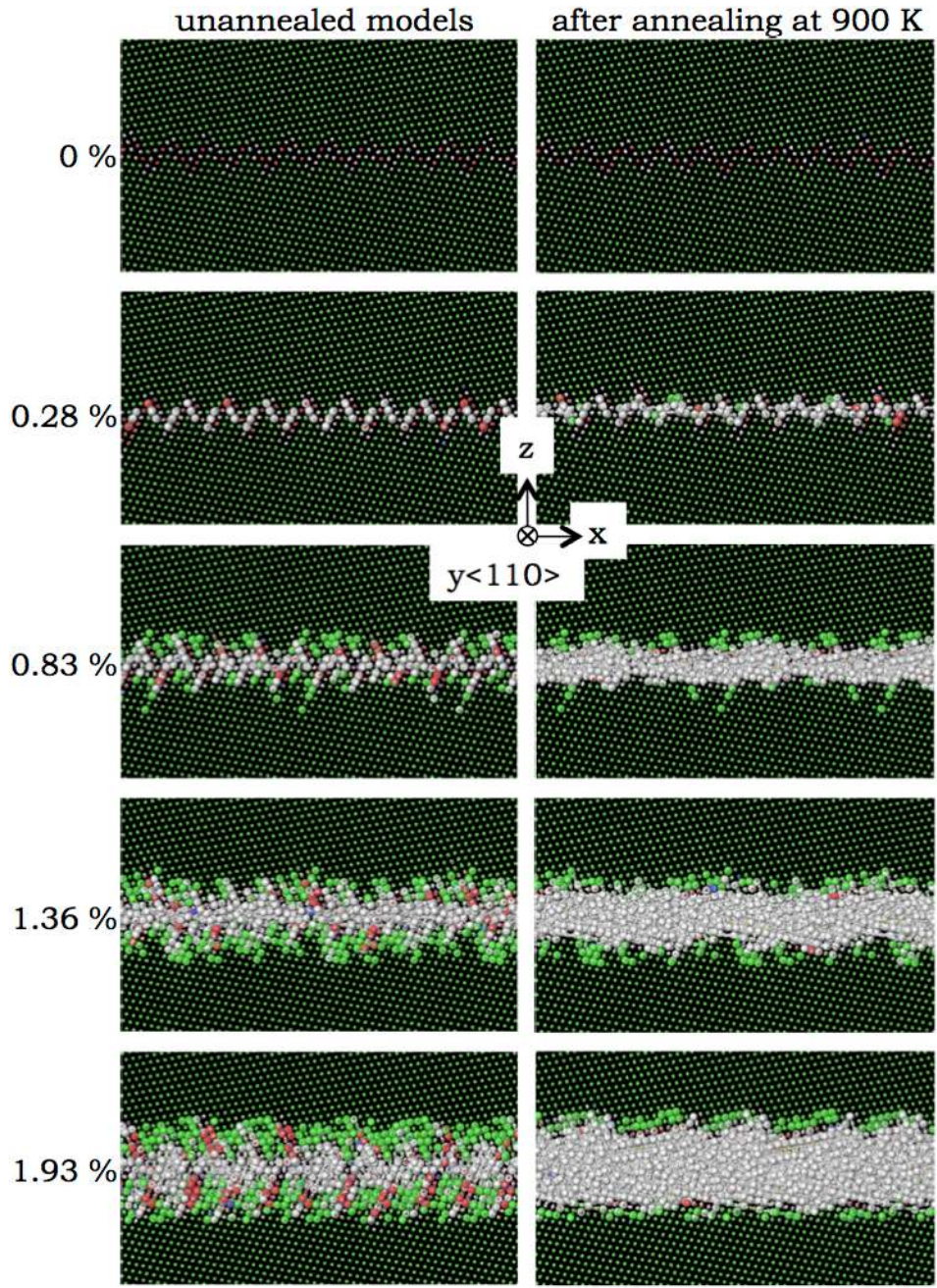

Figure 2. The GB structure for the representative Zr concentrations (just before the emission of the first dislocation). The size of the Cu atoms is artificially decreased to show the positions of the Zr atoms. The atoms are colored according to the common neighbor analysis (CNA) [20,21]. The color-coding is the following: green – fcc, red – hcp, grey – other. Stacking fault segments (shown in red) at the GB indicate the preexisting dislocation embryos.

Annealing the initial model at T=900 K makes the GB structure fully amorphous when the Zr concentration is above 0.83 % (see Fig. 2). In this state, it does not contain the dislocation embryos originally present at the GB in the form of the E structural units [22,23]. In the absence of these dislocation embryos it becomes difficult to nucleate a dislocation from the GB under applied tensile loading, which leads to a very high yield stress. Further addition of Zr makes the GB region wider but does not create new dislocation sources, and the yield stress does not change systematically at larger Zr concentrations.

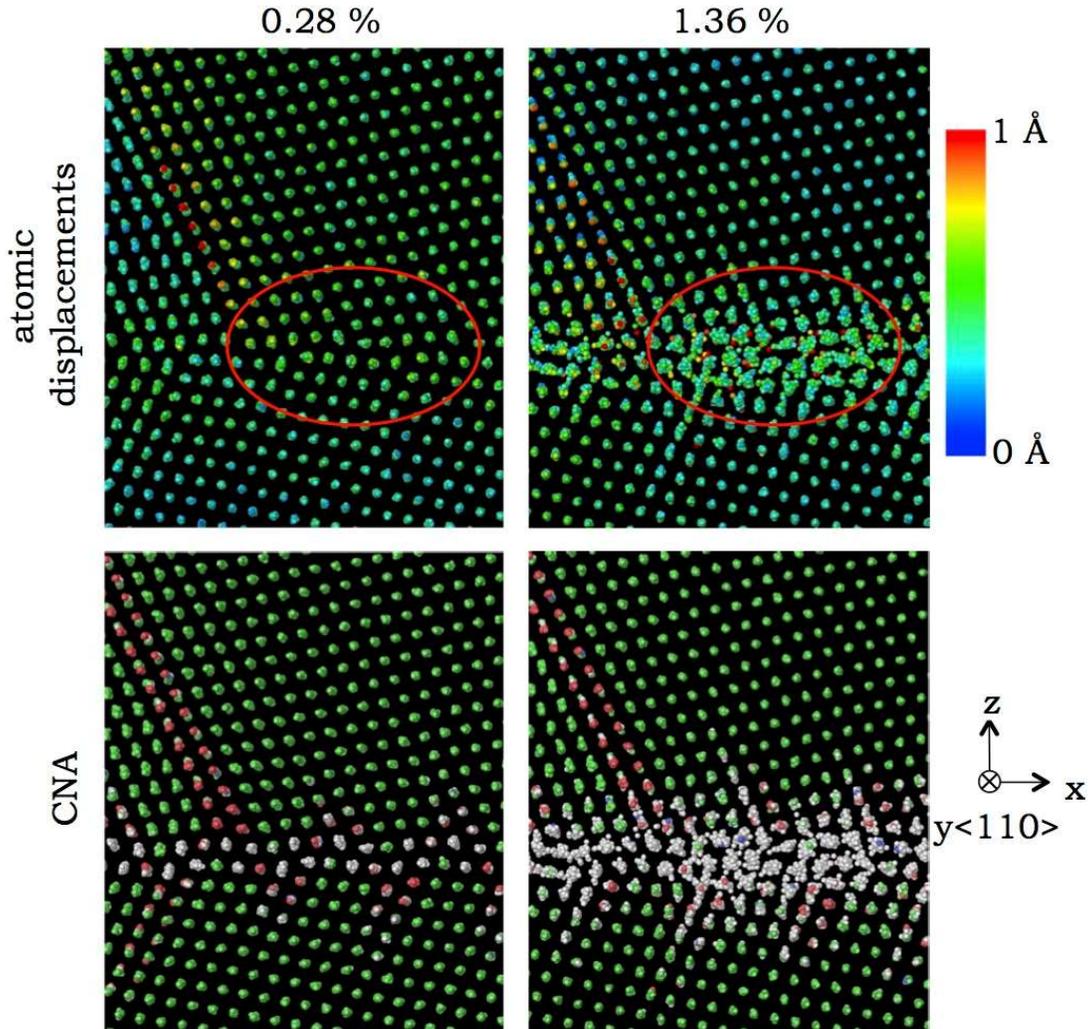

Figure 3. The region of the GB where the first dislocation was emitted for the cases of lower (0.28%) and higher (1.36%) Zr concentrations (unannealed models). The upper images show the atomic displacements caused by the dislocation emission from the GB under applied tensile loading (at T=300K). The size of both solutes and solvents is decreased (and made the same). The lower images are colored according to the CNA [20,21]. The color-coding is the following: green – fcc, red – hcp, grey – other.

Our simulations elucidate the effects of solute-induced disorder of the GB structure on dislocation nucleation. If disorder is complete and the GB does not retain any E units, the yield stress is increased because of the lack of dislocation nucleation sites. Partial disorder can lead to dramatic decreases in the yield stress if some E units remain within the GB, because the atomic shuffling in amorphous regions near the dislocation embryos aids the dislocation emission.

Complete disorder was achieved by annealing at T=900 K for 10 ns, in the case of a Zr-segregated at $\Sigma11(332)[110]$ GB in Cu. Experimental observations may vary because much longer annealing can lead to the formation of Cu-Zr compounds and the interface between these compounds and the bulk Cu can provide new dislocation nucleation sites, as discussed in [10]. The average Zr concentration corresponding to a particular grain boundary composition is also likely to depend on the grain size. Optimal compositions and annealing conditions for coarsening resistance and/or strengthening should be investigated in further studies.

**Acknowledgements:**

This work was supported by the U.S. Department of Energy, Office of Science, Basic Energy Sciences, Materials Science and Engineering Division. The research was performed at Ames Laboratory, which is operated for the U.S. DOE by Iowa State University under contract # DE-AC02-07CH11358